\documentclass[aps,prl, twocolumn,superscriptaddress]{revtex4-2}

\usepackage{dcolumn}
\usepackage{bm}
\usepackage[dvipdfmx]{graphicx} 
\usepackage{color}
\usepackage{braket}

\begin{document}

\title{Role of Gd 5$d$$-$Au 5$d$ hybridization for the magnetic properties of Au-Al-Gd quasicrystal approximants}

\author{Goro Nozue}
\thanks{goro.nozue@uni-wuerzburg.de}
\thanks{Present address: Physikalisches Institut and W\"{u}rzburg-Dresden Cluster of Excellence ctd.qmat, Julius-Maximilians-Universit\"{a}t W\"{u}rzburg, 97074 W\"{u}rzburg, Germany}
\affiliation{Division of Materials Physics, Graduate School of Engineering Science, The University of Osaka, Toyonaka, Osaka  560-8531, Japan}
\affiliation{RIKEN SPring-8 Center, Sayo, Hyogo 679-5148, Japan}

\author{Hidenori Fujiwara}
\affiliation{Division of Materials Physics, Graduate School of Engineering Science, The University of Osaka, Toyonaka, Osaka  560-8531, Japan}
\affiliation{RIKEN SPring-8 Center, Sayo, Hyogo 679-5148, Japan}
\affiliation{Spintronics Research Network Division, Institute for Open and Transdisciplinary Research Initiatives, The University of Osaka, Suita, Osaka 565-0871, Japan}

\author{Satoru Hamamoto}
\affiliation{RIKEN SPring-8 Center, Sayo, Hyogo 679-5148, Japan}

\author{Yuto Torii}
\author{Mai Sakaguchi}
\affiliation{Division of Materials Physics, Graduate School of Engineering Science, The University of Osaka, Toyonaka, Osaka  560-8531, Japan}
\affiliation{RIKEN SPring-8 Center, Sayo, Hyogo 679-5148, Japan}

\author{Sho Nakajima}
\affiliation{Division of Materials Physics, Graduate School of Engineering Science, The University of Osaka, Toyonaka, Osaka  560-8531, Japan}

\author{Takuya D. Nakamura}
\author{Nonoka U. Sakamoto}
\affiliation{Division of Materials Physics, Graduate School of Engineering Science, The University of Osaka, Toyonaka, Osaka  560-8531, Japan}
\affiliation{RIKEN SPring-8 Center, Sayo, Hyogo 679-5148, Japan}

\author{Takayuki Kiss}
\affiliation{Division of Materials Physics, Graduate School of Engineering Science, The University of Osaka, Toyonaka, Osaka  560-8531, Japan}

\author{Masaki Oura}
\affiliation{RIKEN SPring-8 Center, Sayo, Hyogo 679-5148, Japan}

\author{Akira Yasui}
\author{Kotaro Higashi}
\author{Naomi Kawamura}
\affiliation{Japan Synchrotron Radiation Research Institute, Sayo, Hyogo 679-5198, Japan}

\author{Kojiro Mimura}
\affiliation{Department of Physics and Electronics, Graduate School of Engineering, Osaka Metropolitan University, Sakai, Osaka 599-8531, Japan}

\author{Arata Tanaka}
\affiliation{Graduate School of Advanced Science and Engineering, Hiroshima University, Higashi-hiroshima, Hiroshima 739-8530, Japan.}

\author{Asuka Ishikawa}
\thanks{Present address:  Institute of Engineering Innovation, School of Engineering, The University of Tokyo, Tokyo 113-8656, Japan}
\affiliation{Research Institute of Science and Technology, Tokyo University of Science, Tokyo 125-8585, Japan}

\author{Farid Labib}
\affiliation{Research Institute of Science and Technology, Tokyo University of Science, Tokyo 125-8585, Japan}

\author{Shintaro Suzuki}\thanks{Present address: Department of Physics, Aoyama Gakuin University, Sagamihara, Kanagawa 252-5258, Japan}
\author{Takafumi D. Yamamoto}\thanks{Present address: Institute of Liberal Arts and Science, Nagoya University, Nagoya 464-8601, Japan}
\author{Ryuji Tamura}
\affiliation{Department of Materials Science and Technology, Tokyo University of Science, Tokyo 125-8585, Japan}

\author{Akira Sekiyama}
\affiliation{Division of Materials Physics, Graduate School of Engineering Science, The University of Osaka, Toyonaka, Osaka  560-8531, Japan}
\affiliation{RIKEN SPring-8 Center, Sayo, Hyogo 679-5148, Japan}
\affiliation{Spintronics Research Network Division, Institute for Open and Transdisciplinary Research Initiatives, The University of Osaka, Suita, Osaka 565-0871, Japan}
\date{\today}

\begin{abstract}
We have studied the electronic states of magnetic Au-Al-Gd quasicrystal approximants, which exhibit different magnetic ground states at low temperatures, by orbital-selective soft and hard x-ray electron spectroscopies in the paramagnetic phase. Our results reveal that the valence electronic states vary among the Au-Al-Gd approximants with different magnetic ground states, whereas the Gd 4$f$ electrons are localized in all of them. In particular, we find that the Gd 5$d$ states are hybridized with the Au 5$d$ states around the Fermi level, providing information about a local pathway for the intersite exchange interaction between the localized Gd 4$f$ electrons. This hybridization strength varies among these samples. Our findings bring an orbital-resolved microscopic picture to the magnetism of rare-earth-based quasicrystals and approximants.

\end{abstract}

\maketitle

\clearpage

\section{Introduction}
Quasicrystals (QCs) are alloys that possess long-range ordered atomic structures with crystallographically forbidden rotational symmetries \cite{a}. Understanding their quasiperiodic structures has been a central issue since the discovery of QCs. Recently, strongly correlated electronic states in rare-earth-based QCs and their approximants (ACs) have also attracted considerable attention, following the discovery of exotic quantum states: long-range magnetic order \cite{b,c,d}, non-Fermi-liquid behavior \cite{e}, and superconductivity \cite{f}. Among these fascinating electronic states, the magnetism of QCs and ACs has been extensively studied \cite{i, ah}.\par

\begin{figure}
	\begin{center}
	\includegraphics[keepaspectratio,,scale=0.7]{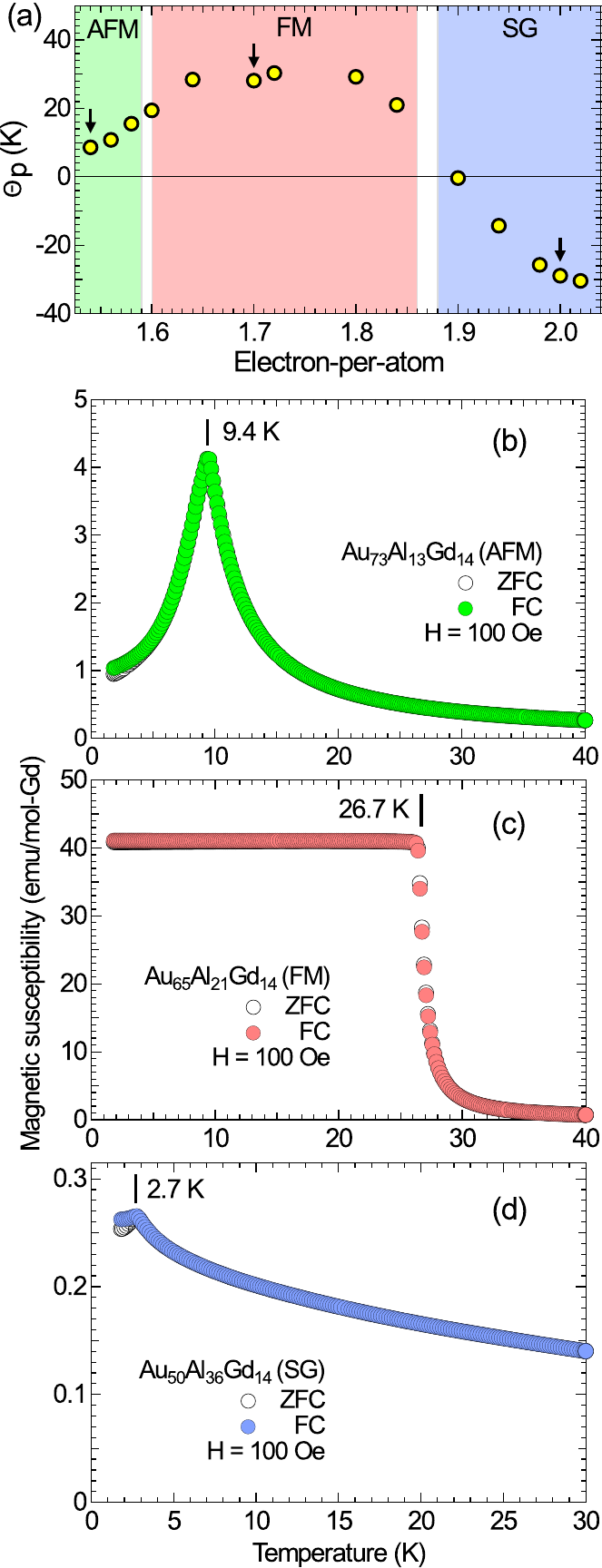}
	\caption{(a) $e/a$-ratio dependence of the Curie-Weiss temperature $\Theta_p$ of the Au-Al-Gd ACs \cite{g,h,i,ag}. The arrows show the $e/a$-ratio values of each Au-Al-Gd AC employed in this study. (b), (c), (d) Temperature dependence of the field cooling (FC) and zero-field cooling (ZFC) magnetic susceptibility of $\rm{Au_{73}}$$\rm{Al_{13}}$$\rm{Gd_{14}}$ (AFM), $\rm{Au_{65}}$$\rm{Al_{21}}$$\rm{Gd_{14}}$ (FM), and $\rm{Au_{50}}$$\rm{Al_{36}}$$\rm{Gd_{14}}$ (SG), respectively. The solid lines indicate the magnetic transition temperatures. }
	\label{Fig1}
	\end{center}
\end{figure}

From the previous investigations of macroscopic magnetic properties of rare-earth-based ACs, it has been revealed that certain rare-earth-based ACs show antiferromagnetic (AFM), ferromagnetic (FM), and spin-glass (SG) ground states with varying electron-per-atom ($e/a$) ratios \cite{ah, i}. Here, the $e/a$ ratio stands for the average number of valence electrons per atom. As an example, Figure \ref{Fig1}(a) shows the paramagnetic Curie-Weiss temperature ($\Theta_p$) as a function of the $e/a$ ratio in the case of Au–Al–Gd ACs \cite{g,h,i,ag}, together with their corresponding magnetic ground states. In Fig.~\ref{Fig1}(a), the value and sign of $\Theta_p$ depend mainly on the $e/a$ ratio, and consequently the magnetic ground state can be grouped into AFM, FM, and SG phases, according to the $e/a$ ratio. Note that the $e/a$ ratio is tuned by changing the Au/Al composition ratio while keeping the Gd concentration constant in Fig.~\ref{Fig1}(a). Importantly, by adjusting the $e/a$ ratio, the FM and AFM orders have also been realized in rare-earth-based QCs \cite{b,c,d}. On the other hand, the systematic understanding of the discovered magnetisms of rare-earth-based QCs and ACs, including the modulation of magnetic ground states with the $e/a$ ratio, remains insufficient, despite extensive studies to date \cite{ag, j,k,l,m}.\par

Valence electronic states, including 4$f$ orbitals, are important keys to understanding the magnetisms of rare-earth-based QCs and ACs. This is because, in general, the magnetic properties of rare-earth-based compounds are basically explained by the Ruderman-Kittel-Kasuya-Yosida (RKKY) interaction, which is an indirect exchange interaction between localized 4$f$ orbitals mediated by itinerant valence bands. The Gd-based QCs and ACs can provide a good platform for studying the relationship between valence electronic states and magnetic properties since the single-ion magnetic anisotropy due to the spin-orbit coupling and the crystalline electric field is negligible for a localized Gd$\rm{^{3+}}$ ion with the $4f^7$ configuration. The previous theoretical calculation has qualitatively explained the modulation of magnetic ground states in the Au-Al-Gd ACs with the $e/a$ ratio, using a minimal magnetic model including only the RKKY interaction \cite{j}. In that study, the variation in the Au concentration was approximated by changes in the Fermi wavenumber ($k_F$). Therefore, it remains unclear how valence electronic states are related to the magnetic properties of rare-earth-based ACs, and which orbitals contribute to them. This motivates us to experimentally investigate the orbital-resolved valence electronic states of the Au-Al-Gd ACs.\par

In this paper, we report the valence electronic states of the Au-Al-Gd ACs, which show different magnetic ground states at low temperatures, in the paramagnetic phase. To probe the orbital-resolved valence electronic states including the localized Gd 4$f$ orbital, we have performed soft and hard x-ray electron spectroscopies, combining hard x-ray photoemission spectroscopy (HAXPES), Gd $M$$\rm{_{4,5}}$-edge x-ray absorption spectroscopy (XAS), Gd 3$d$-edge resonant photoemission spectroscopy (RPES), and high-energy-resolution fluorescence-detected XAS (HERFD-XAS). Our results show that the valence electronic states are systematically different in the Au-Al-Gd ACs, while the Gd 4$f$ electrons are localized in all samples. Moreover, we find hybridization between the Gd 5$d$ and Au 5$d$ states around the Fermi level ($E_F$), offering a microscopic picture of the intersite exchange interaction between the localized Gd 4$f$ electrons. The effective strength of this hybridization is weaker in the spin-glass phase sample than that in the antiferromagnet and ferromagnet samples.\par

\section{Experimental Details}
 Polycrystalline samples of antiferromagnet $\rm{Au_{73}}$$\rm{Al_{13}}$$\rm{Gd_{14}}$ (denoted as AFM), ferromagnet $\rm{Au_{65}}$$\rm{Al_{21}}$$\rm{Gd_{14}}$ (denoted as FM), and spin-glass compound $\rm{Au_{50}}$$\rm{Al_{36}}$$\rm{Gd_{14}}$ (denoted as SG) were synthesized by arc-melting high-purity Au, Al, and Gd elements and subsequently annealed at 1073 K for 50 h under an Ar atmosphere \cite{g,h}. \color{black}Each $e/a$-ratio value of the Au-Al-Gd ACs is 1.54, 1.70, and 2.00 for $\rm{Au_{73}Al_{13}Gd_{14}}$ (AFM), $\rm{Au_{65}Al_{21}Gd_{14}}$ (FM), and $\rm{Au_{50}Al_{36}Gd_{14}}$ (SG), respectively (displayed in Fig.~\ref{Fig1}(a)). The macroscopic magnetic ground states were confirmed by the temperature dependence of magnetic susceptibility measurements shown in Figs.~\ref{Fig1}(b), (c), and (d), demonstrating the N\'{e}el temperature of 9.4 K for $\rm{Au_{73}}$$\rm{Al_{13}}$$\rm{Gd_{14}}$, the Curie temperature of 26.7 K for $\rm{Au_{65}}$$\rm{Al_{21}}$$\rm{Gd_{14}}$, and the spin freezing temperature of 2.7 K for $\rm{Au_{50}}$$\rm{Al_{36}}$$\rm{Gd_{14}}$.\par

The core-level and valence-band HAXPES was carried out at BL09XU of SPring-8 using a Scienta Omicron R4000 hemispherical analyzer \cite{n}. The energy of the incident hard x-ray was set to 7.2 keV. The overall energy resolution was 150 meV for the HAXPES. The Gd $M$$\rm{_{4,5}}$-edge XAS and Gd 3$d$-edge RPES were performed at BL17SU of SPring-8 \cite{o}. The XAS spectra were obtained in a total-electron-yield mode. The RPES valence-band spectra were recorded using a Scienta Omicron EW4000 electron analyzer. The overall energy resolution was set to 200 meV with the photon energy of $\sim$ 1184.8 eV corresponding to the Gd $M_5$ edge. The clean surfaces were obtained by $\it in$-$\it situ$ fracturing for the soft and hard x-ray spectroscopies. The temperature was set to 40 K, at which all samples were in a paramagnetic phase, for both measurements. \par

The Gd and Au $L_3$-edge HERFD-XAS was conducted at BL39XU of SPring-8 \cite{p,q}. The $L$$\alpha_1$ fluorescence of Gd and Au was selected by the Si(333) and the Ge(555) crystal-analyzer spectrometers, respectively. The selected fluorescence was recorded by the two-dimensional pixel detector PILATUS \cite{r}. The temperature for the HERFD-XAS was set to 40 K.\par

The ion-model simulations for the Gd 3$d$ core-level HAXPES and Gd $M_{4,5}$-edge XAS spectra were calculated by the XTLS ver.~9.0 program, including the full multiplet structures \cite{s}. The atomic parameters for the Gd$\rm{^{3+}}$ ion, like the 3$d$-4$f$ and 4$f$-4$f$ Coulomb and exchange interactions (Slater integrals) and the 3$d$ and 4$f$ spin-orbit coupling, were calculated by Cowan's code based on the Hartree-Fock method \cite{t}. The 3$d$-4$f$ and 4$f$-4$f$ Slater integrals (3$d$ and 4$f$ spin-orbit coupling) were reduced to 85\% and 87\% (99\% and 100\%) to reproduce the experimental spectra with final-state multiplet structure, respectively.\par

\section{Results and discussion}

\begin{figure}
         
	\begin{center}
	\includegraphics[keepaspectratio,,scale=0.65]{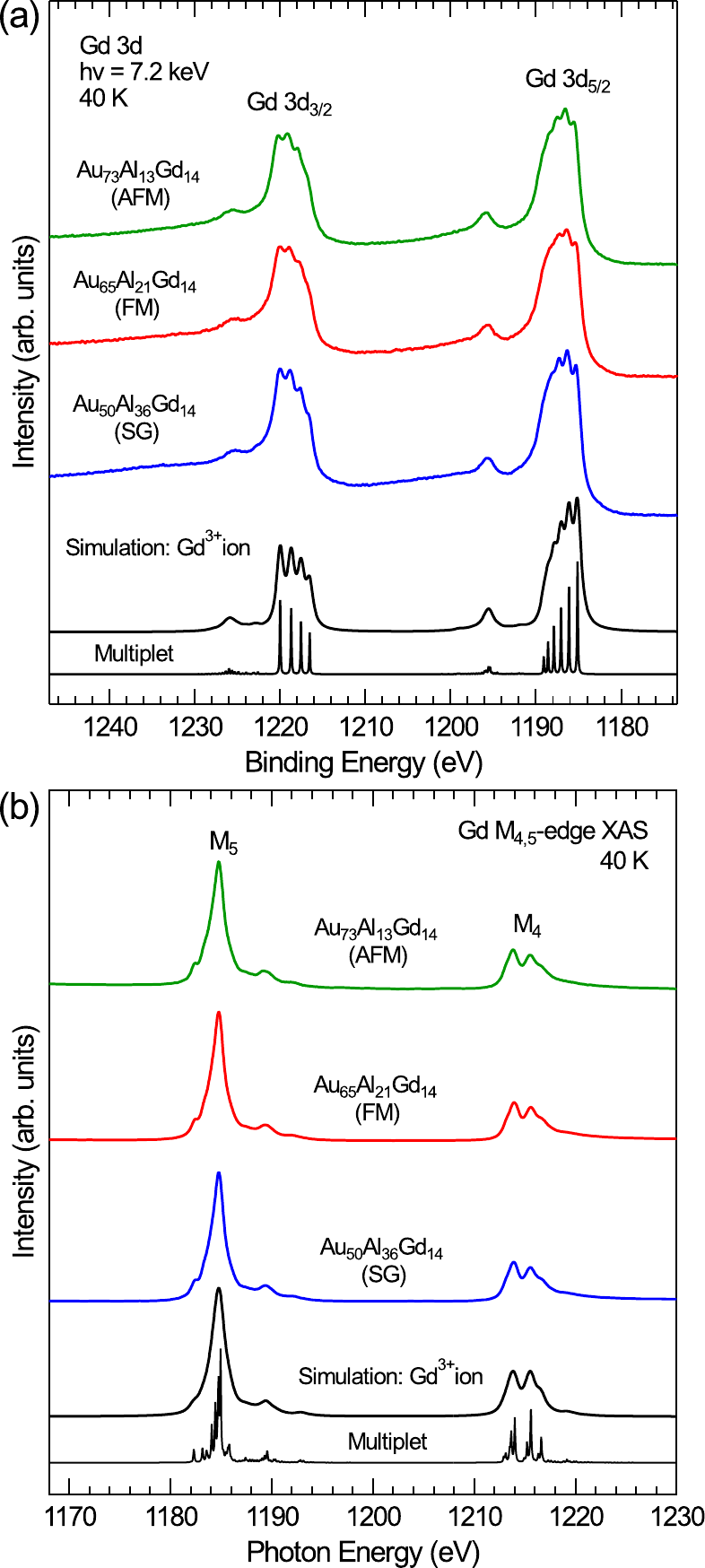}
	\caption{(a) Gd 3$d$ core-level HAXPES and (b) Gd $M_{4,5}$-edge XAS spectra of $\rm{Au_{73}}\rm{Al_{13}}\rm{Gd_{14}}$ (AFM), $\rm{Au_{65}}\rm{Al_{21}}\rm{Gd_{14}}$ (FM), and $\rm{Au_{50}}\rm{Al_{36}}\rm{Gd_{14}}$ (SG), compared with the simulated spectra for the Gd$^{3+}$ ion and the corresponding final-state multiplet structures, 3$d^9$4$f^7$ for (a) and 3$d^9$4$f^8$ for (b).}
	\label{Fig2}
	\end{center}
\end{figure}

\begin{figure*}
	\begin{center}
	\includegraphics[keepaspectratio,,scale=0.8]{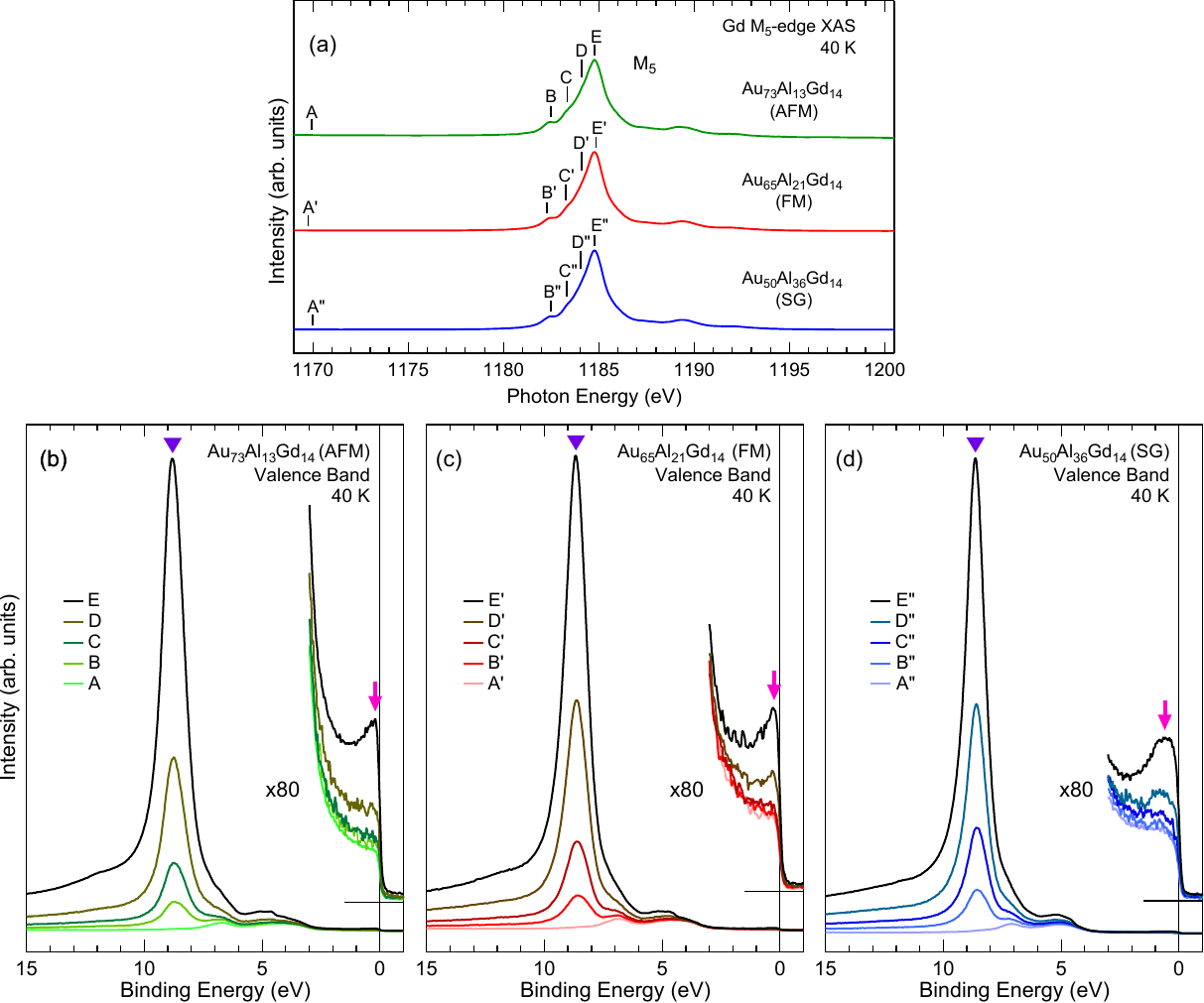}
	\caption{(a) Gd $M_5$-edge XAS spectra of $\rm{Au_{73}}$$\rm{Al_{13}}$$\rm{Gd_{14}}$ (AFM), $\rm{Au_{65}}$$\rm{Al_{21}}$$\rm{Gd_{14}}$ (FM), and $\rm{Au_{50}}$$\rm{Al_{36}}$$\rm{Gd_{14}}$ (SG). The vertical bars indicate the selected photon energy for the Gd 3$d$-edge RPES of the Au-Al-Gd ACs. (b), (c), (d) Gd 3$d$-edge RPES spectra of  $\rm{Au_{73}}$$\rm{Al_{13}}$$\rm{Gd_{14}}$ (AFM), $\rm{Au_{65}}$$\rm{Al_{21}}$$\rm{Gd_{14}}$ (FM), and $\rm{Au_{50}}$$\rm{Al_{36}}$$\rm{Gd_{14}}$ (SG), respectively. The triangles indicate the 4$f^6$ final-state multiplet structures. The downward arrows show the resonant peak structure around $E_F$.}
	\label{Fig3}
	\end{center}
\end{figure*}

\subsection{Gd 3$d$ core-level HAXPES and Gd $M$$\rm{_{4,5}}$-edge XAS} 
We start with a discussion of the Gd 4$f$ electronic states in the Au-Al-Gd ACs, as probed by the Gd 3$d$ core-level HAXPES, which is sensitive to element-selective local electronic states. Figure \ref{Fig2}(a) shows the Gd 3$d$ core-level HAXPES spectra of $\rm{Au_{73}}$$\rm{Al_{13}}$$\rm{Gd_{14}}$ (AFM), $\rm{Au_{65}}$$\rm{Al_{21}}$$\rm{Gd_{14}}$ (FM), and $\rm{Au_{50}}$$\rm{Al_{36}}$$\rm{Gd_{14}}$ (SG). In these three spectra, the two main peaks are observed at around 1220 eV and 1187 eV, which are ascribed to the Gd 3$d$$\rm{_{3/2}}$ and 3$d$$\rm{_{5/2}}$ core-level excitations, respectively. Moreover, there are several additional peaks in both main peaks of all samples. The line shapes of these spectra are mutually similar to each other. Figure \ref{Fig2}(a) also displays the simulated spectrum for the Gd$\rm^{3+}$ ion and the 3$d^9$4$f^7$ final-state multiplet structures. The simulated spectrum reproduces the overall spectral shape of the experimental Gd 3$d$ core-level spectra. In particular, the several peaks around the main peaks are explained by the multiplet structures \cite{af}. Therefore, the Gd 4$f$ electrons are well localized in all Au-Al-Gd ACs. Note that the Gd 4$d$ core-level HAXPES spectra of the Au-Al-Gd ACs are also explained by the calculated spectrum for the Gd$\rm^{3+}$ ion (see Appendices). \par

The Gd $M_{4,5}$-edge XAS is also sensitive to the local Gd 4$f$ electronic states owing to strong Coulomb attractive interactions between the 3$d$ core hole and local 4$f$ electrons. Figure \ref{Fig2}(b) displays the Gd $M_{4,5}$-edge XAS spectra of the Au-Al-Gd ACs together with the Gd$^{3+}$ ion-model simulation spectrum and the simulated 3$d^9$4$f^8$ final-state multiplet structures. The simulated spectrum reproduces the line shape of the experimental main peaks centered at $\sim$ 1185 eV and $\sim$ 1215 eV, ascribed to the $M_5$ and $M_4$ edges, respectively. In addition, the several peaks and hump structures at both main peaks are understood as the 3$d^9$4$f^8$ final-state multiplet structures. The localized Gd 4$f$ electronic states in the Au-Al-Gd ACs are also confirmed by the Gd $M$$\rm{_{4,5}}$-edge XAS. This is consistent with the results of Gd 3$d$ and 4$d$ core-level HAXPES.\par

\begin{figure}
	\begin{center}
	\includegraphics[keepaspectratio,,scale=0.49]{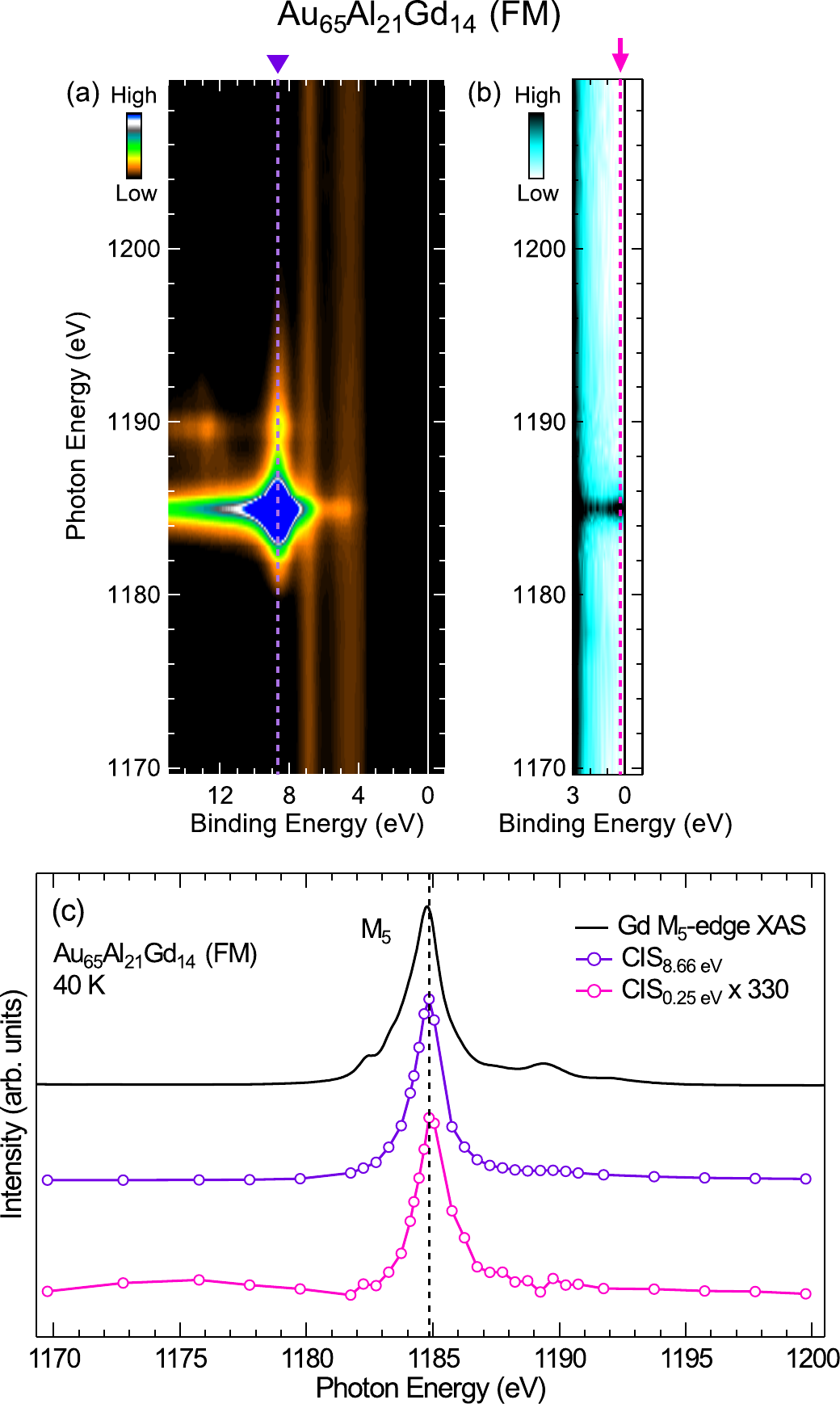}
	\caption{(a) Intensity map for the Gd 3$d$-edge RPES spectra of $\rm{Au_{65}}$$\rm{Al_{21}}$$\rm{Gd_{14}}$ (FM). The triangle shows the peak binding energy of the 4$f^6$ final-state multiplet structures. (b) Enlarged view of the intensity map for the  Gd 3$d$-edge RPES spectra of $\rm{Au_{65}}$$\rm{Al_{21}}$$\rm{Gd_{14}}$ (FM) around $E_F$. The downward arrow indicates the binding energy of the resonant peak structure in the vicinity of $E_F$. In (a) and (b), the spectral intensity has been linearly interpolated with respect to the photon energy. (c) CIS spectra at 8.66 and 0.25 eV in the Gd 3$d$-edge RPES spectra of $\rm{Au_{65}}$$\rm{Al_{21}}$$\rm{Gd_{14}}$ (FM) in comparison with the Gd $M_5$-edge XAS spectrum.}
	\label{Fig4}
	\end{center}
\end{figure}

\subsection{Gd 3$d$-edge RPES}
To directly visualize the Gd 4$f$ electronic states as well as the Gd 5$d$ electronic states, we have performed the Gd 3$d$-edge RPES on the Au-Al-Gd ACs. Figure \ref{Fig3}(a) presents the Gd $M$$\rm{_5}$-edge XAS spectra of the Au-Al-Gd ACs, in which the photon energies employed for the Gd 3$d$-edge RPES are indicated by the vertical bars and labeled. Figures \ref{Fig3}(b), (c), and (d) display the Gd 3$d$-edge RPES spectra of $\rm{Au_{73}}$$\rm{Al_{13}}$$\rm{Gd_{14}}$ (AFM), $\rm{Au_{65}}$$\rm{Al_{21}}$$\rm{Gd_{14}}$ (FM), and $\rm{Au_{50}}$$\rm{Al_{36}}$$\rm{Gd_{14}}$ (SG), respectively. All valence-band RPES spectra are normalized by the photon flux. On going from the photon energy A (A', A": $h$$\nu$$\sim$ 1170 eV) to E (E', E": $h$$\nu$$\sim$ 1184.8 eV), the significant resonance enhancement of the peak structure is observed at around the binding energy of 8.7 eV in Figs.~\ref{Fig3}(b), (c), and (d). This peak structure is explained by the 4$f^6$ final-state multiplet structures of the Gd$\rm^{3+}$ state \cite{u}, indicating the dominant localized Gd 4$f$ electronic states in all samples. On the other hand, the slight evolution of the peak structure around $E_F$, which is not explained by the 4$f^6$ final-state multiplet structures, is also confirmed in Figs.~\ref{Fig3}(b), (c), and (d). It should be emphasized that these peak structures are not due to the effect of Auger emission because the binding energy of these peak structures remains constant as the photon energy is changed.

To clarify the origin of the resonance enhancement near $E_F$, the detailed Gd 3$d$-edge RPES has also been carried out in $\rm{Au_{65}}$$\rm{Al_{21}}$$\rm{Gd_{14}}$ (FM). Figure \ref{Fig4}(a) shows the intensity map for the RPES spectra of $\rm{Au_{65}}$$\rm{Al_{21}}$$\rm{Gd_{14}}$ (FM) across the $M_5$ edge as a function of the binding energy and photon energy. The prominent intensity is observed at the binding energy of 8.66 eV with the photon energy of  $\sim$ 1184.8 eV, corresponding to the $M_5$ edge. This structure is ascribed to the 4$f^6$ final-state multiplet structures discussed in Fig.~\ref{Fig3}(c). Additionally, there is a noticeable intensity at around the binding energy of 13 eV with the photon energy of $\sim$ 1189 eV in  Fig.~\ref{Fig4}(a). These structures are confirmed in the Gd 3$d$-edge RPES spectra of $\rm{Au_{65}}$$\rm{Al_{21}}$$\rm{Gd_{14}}$ (FM) shown in Appendices. Based on the previous studies \cite{w,x}, this feature can be interpreted as the effect of the spin-flip transition in the Gd 3$d$-edge RPES. Figure \ref{Fig4}(b) displays the enlarged view of the intensity map around $E_F$ with a different color scale from Fig.~\ref{Fig4}(a).  The low intensity corresponding to the resonance peak structure near $E_F$ is confirmed at around the binding energy of 0.25 eV with the photon energy of $\sim$ 1184.8 eV.\par

For the detailed discussion on the resonance component around $E_F$, we show the photon energy dependence of the photoelectron intensity at binding energies of 8.66 eV (the 4$f^6$ final-state multiplet structures) and 0.25 eV (the resonance peak structure around $E_F$) in the Gd 3$d$-edge RPES spectra of $\rm{Au_{65}}$$\rm{Al_{21}}$$\rm{Gd_{14}}$ (FM) in Fig.~\ref{Fig4}(c). These photon-energy-dependent intensity profiles are called constant-initial-state (CIS) spectra. The Gd $M$$\rm{_{5}}$-edge XAS spectrum of $\rm{Au_{65}}$$\rm{Al_{21}}$$\rm{Gd_{14}}$ (FM) is also displayed in Fig.~\ref{Fig4}(c) as a reference. The spectral intensity of the CIS at 0.25 eV is scaled by a factor of 330 for the line-shape comparison. Both CIS spectra show the peak structure at the photon energy of 1184.85 eV corresponding to the $M_5$ edge. If the resonance component near $E_F$ is ascribed to the Gd 4$f$ orbital from the Gd$^{2+}$ state owing to the effect of hybridization with conduction electrons, the peak photon energy of the CIS spectrum at 0.25 eV should be different from that at 8.66 eV. This is because the values of the 3$d$-4$f$ and 4$f$-4$f$ Slater integrals vary between the Gd$^{3+}$ and Gd$^{2+}$ states due to the different number of 4$f$ electrons. Indeed, mutually different  $h$$\nu$-dependent resonance enhancements between the trivalent-like and divalent-like 4$f$ components have been observed for a Nd-based compound \cite{az}. In addition,  the simulated spectra for the Gd$\rm^{3+}$ ion reproduce the spectral shape of experimental Gd 3$d$ and 4$d$ core-level HAXPES and Gd $M$$\rm{_{4,5}}$-edge XAS spectra. Therefore, the resonance component around $E_F$ does not originate from the Gd 4$f$ contribution.\par

\begin{figure}[t]
	\begin{center}
	\includegraphics[keepaspectratio,,scale=0.49]{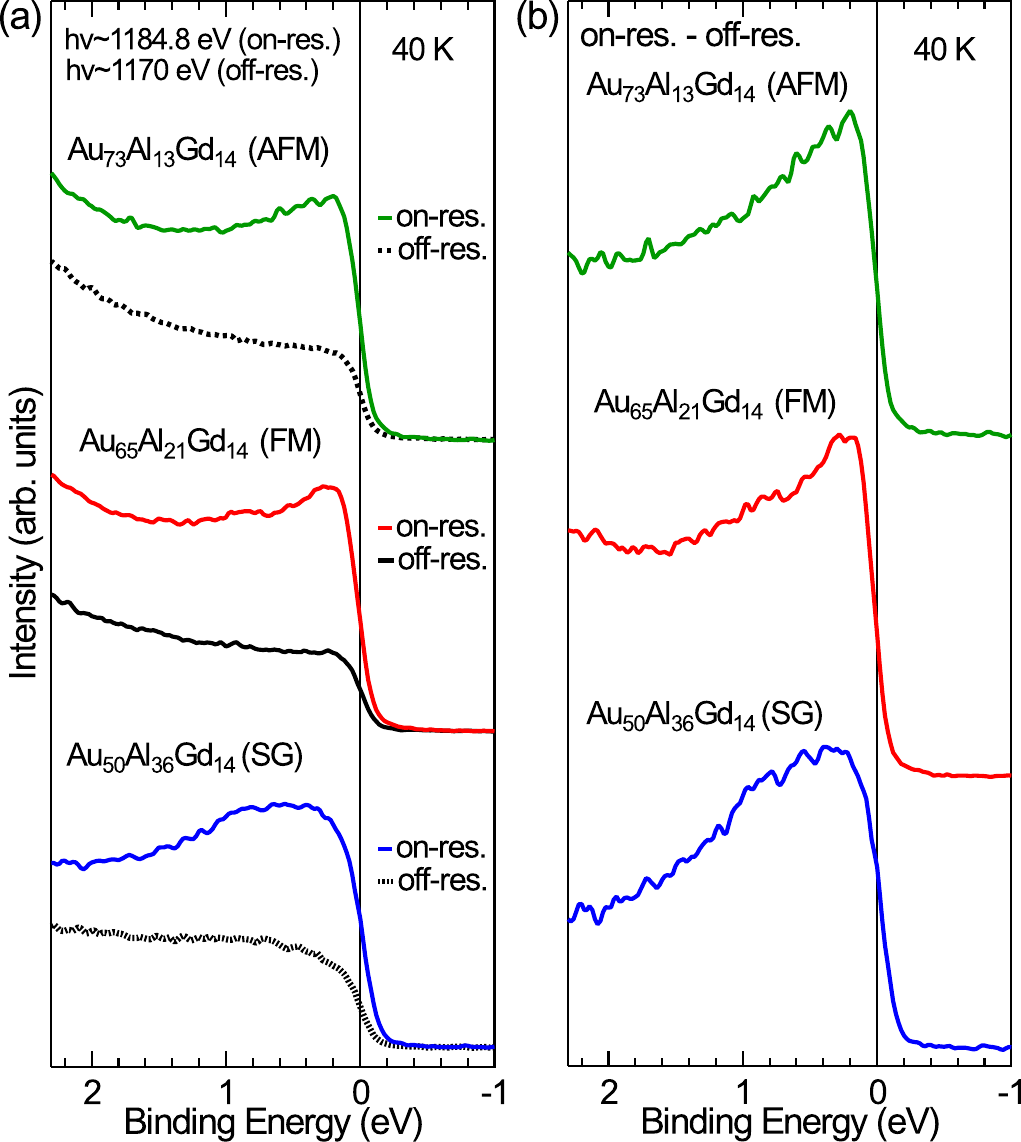}
	\caption{(a) Enlarged view of the on-res.~and off-res.~spectra of  $\rm{Au_{73}}$$\rm{Al_{13}}$$\rm{Gd_{14}}$ (AFM), $\rm{Au_{65}}$$\rm{Al_{21}}$$\rm{Gd_{14}}$ (FM), and $\rm{Au_{50}}$$\rm{Al_{36}}$$\rm{Gd_{14}}$ (SG) around $E_F$. (b) Difference spectra (on-res.~spectrum $-$ off-res.~spectrum) of $\rm{Au_{73}}$$\rm{Al_{13}}$$\rm{Gd_{14}}$ (AFM), $\rm{Au_{65}}$$\rm{Al_{21}}$$\rm{Gd_{14}}$ (FM), and $\rm{Au_{50}}$$\rm{Al_{36}}$$\rm{Gd_{14}}$ (SG).}
	\label{Fig5}
	\end{center}
\end{figure}

In the Gd 3$d$-edge RPES, the resonance enhancement of the Gd 4$f$ orbital takes place due to the quantum interference effect between the two different processes with the same initial and final states: the direct photoemission process $(\ket{3d^{10}4f^{7}}$$+$$h\nu\rightarrow\ket{3d^{10}4f^{6}}+e^-)$ and the 4$f$ Auger process $(\ket{3d^{10}4f^{7}}+h\nu\rightarrow\ket{3d^{9}4f^8}\rightarrow\ket{3d^{10}4f^{6}}+e^-)$. In addition to the 4$f$ Auger process, the 5$d$ Auger process also occurs in the Gd 3$d$-edge RPES. Since the initial and final states of the direct photoemission process $(\ket{3d^{10}4f^{7}5d^{n}}+h\nu\rightarrow\ket{3d^{10}4f^{7}5d^{n-1}}+e^-)$  are identical with those of the 5$d$ Auger process $(\ket{3d^{10}4f^{7}5d^{n}}+h\nu\rightarrow\ket{3d^{9}4f^{8}5d^{n}}\rightarrow\ket{3d^{10}4f^{7}5d^{n-1}}+e^- )$, the Gd 5$d$ states can be resonantly enhanced in the Gd 3$d$-edge RPES. Indeed, the resonance enhancement of  La, Yb, and Tb 5$d$ states has been observed in the La, Yb, and Tb 3$d$-edge RPES for La$\rm{Cu_2}$$\rm{Ge_2}$ \cite{z} and La$\rm{Ni_2}$$\rm{Ge_2}$ \ \cite{aa}, Yb$\rm{Rh_2}$$\rm{Si_2}$ \cite{y}, and the magnetic Au-Al-Tb ACs \cite{ab}, respectively. A peak photon energy in the CIS spectrum of the 5$d$ orbital coincides with that of the 4$f$ orbitals because both 5$d$ and 4$f$ Auger processes have the same intermediate states. In our results, the CIS spectra at 0.25 eV and 8.66 eV show the peak structure at the same photon energy in Fig. \ref{Fig4}(c). Based on that, we conclude that the resonance enhancement around $E_F$ is due to the Gd 5$d$-orbital contribution. \par

To compare the Gd 5$d$ electronic states in the Au-Al-Gd ACs, we show the on-resonance (on-res.) spectra at the photon energy of $\sim$ 1184.8 eV and the off-resonance (off-res.) spectra at the photon energy of $\sim$ 1170 eV of the Au-Al-Gd ACs in Fig.~\ref{Fig5}(a). In all samples, the line shape of the on-res.~spectra is different from that of the off-res.~spectra due to the resonance enhancement of the Gd 5$d$ states. Figure \ref{Fig5}(b) displays the difference spectra obtained from the on-res.~and off-res.~spectra of each sample. These difference spectra dominantly reflect the Gd 5$d$ electronic states. The difference spectrum of $\rm{Au_{50}}$$\rm{Al_{36}}$$\rm{Gd_{14}}$ (SG) shows the broader peak compared with that of  $\rm{Au_{73}}$$\rm{Al_{13}}$$\rm{Gd_{14}}$ (AFM) and $\rm{Au_{65}}$$\rm{Al_{21}}$$\rm{Gd_{14}}$ (FM). Moreover, the peak structure near $E_F$ is relatively steep for $\rm{Au_{73}Al_{13}Gd_{14}}$ (AFM), while the peak maximum around $E_F$ shows the flat spectral shape for $\rm{Au_{65}}$$\rm{Al_{21}Gd_{14}}$ (FM). In contrast to the highly localized Gd 4$f$ electronic states in all samples, the Gd 5$d$ electronic states are different among the Au-Al-Gd ACs. \par

\begin{figure}[t]
	\begin{center}
	\includegraphics[keepaspectratio,,scale=0.54]{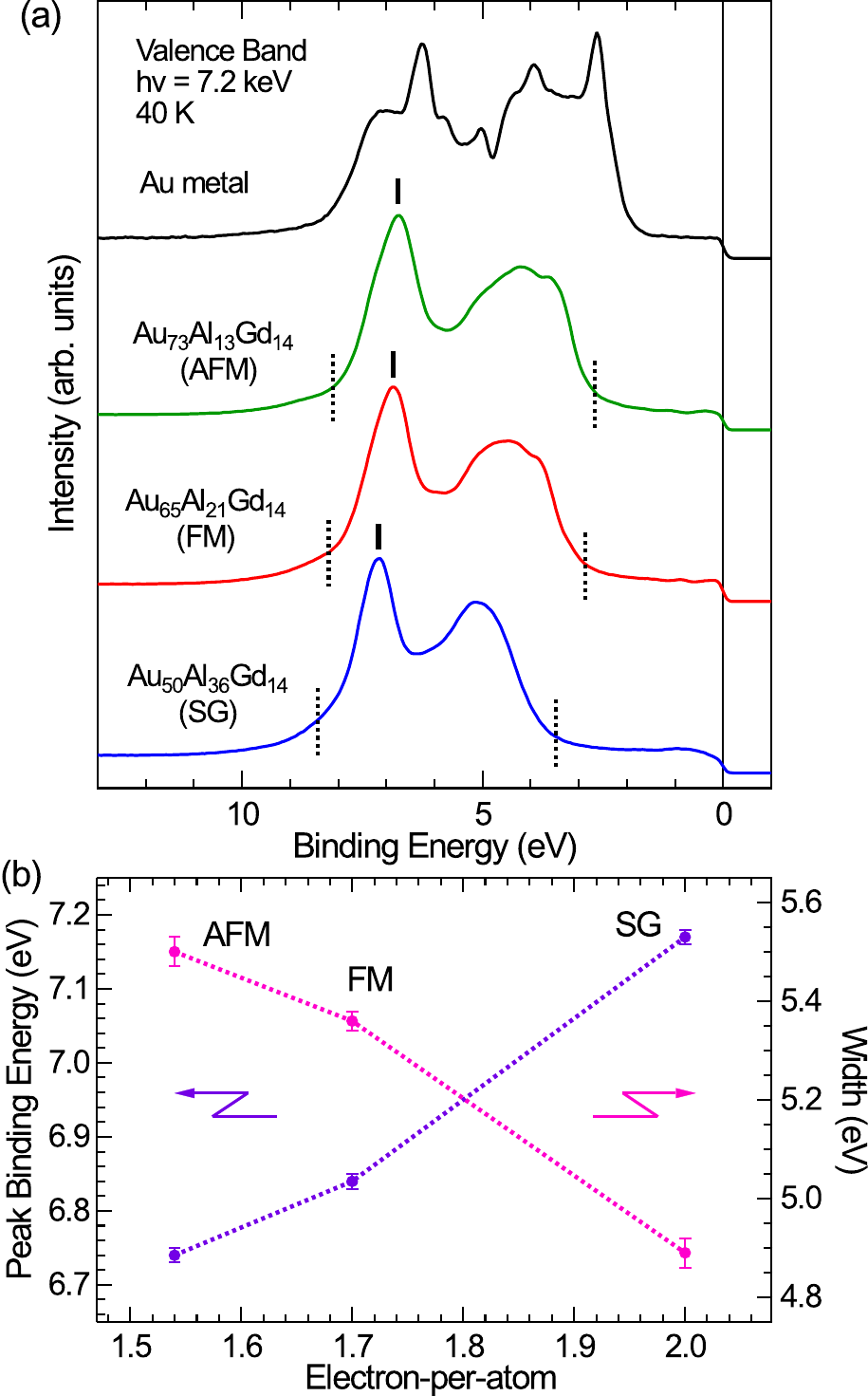}
	\caption{(a) Valence-band HAXPES spectra of $\rm{Au_{73}}$$\rm{Al_{13}}$$\rm{Gd_{14}}$ (AFM), $\rm{Au_{65}}$$\rm{Al_{21}}$$\rm{Gd_{14}}$ (FM), $\rm{Au_{50}}$$\rm{Al_{36}}$$\rm{Gd_{14}}$ (SG) and Au metal. The solid lines indicate the peak binding energy of the Au 5$d$ states. The dashed lines show the bandwidth of the Au 5$d$ bands. (b) Value of peak binding energy and bandwidth of the Au 5$d$ band in the Au-Al-Gd ACs as a function of the $e/a$ ratio. The Au 5$d$ bandwidth has been estimated by linearly extrapolating the tails of the 5$d$ band.}
	\label{Fig6}
	\end{center}
\end{figure}

\begin{figure}[t]
	\begin{center}
	\includegraphics[keepaspectratio,,scale=0.43]{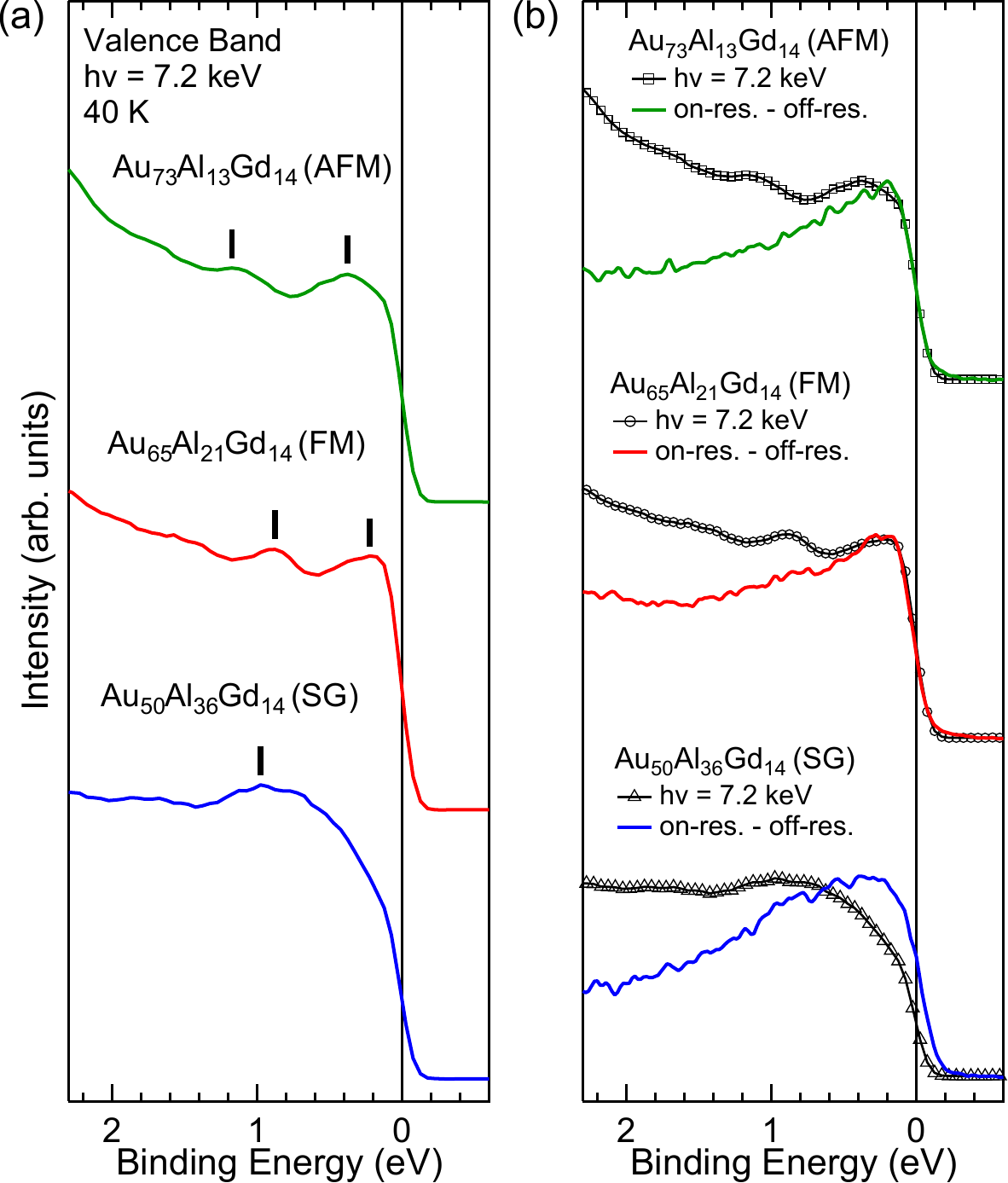}
	\caption{(a) Enlarged view of the valence-band HAXPES spectra of $\rm{Au_{73}}$$\rm{Al_{13}}$$\rm{Gd_{14}}$ (AFM), $\rm{Au_{65}}$$\rm{Al_{21}}$$\rm{Gd_{14}}$ (FM), and $\rm{Au_{50}}$$\rm{Al_{36}}$$\rm{Gd_{14}}$ (SG) around $E_F$. The solid lines indicate the peaks in each spectrum. (b) Comparison of the difference spectra from the Gd 3$d$-edge RPES and the valence-band HAXPES spectra of $\rm{Au_{73}}$$\rm{Al_{13}}$$\rm{Gd_{14}}$ (AFM), $\rm{Au_{65}}$$\rm{Al_{21}}$$\rm{Gd_{14}}$ (FM), and $\rm{Au_{50}}$$\rm{Al_{36}}$$\rm{Gd_{14}}$ (SG).}
	\label{Fig7}
	\end{center}
\end{figure}

\begin{figure*}
	\begin{center}
	\includegraphics[keepaspectratio,,scale=0.88]{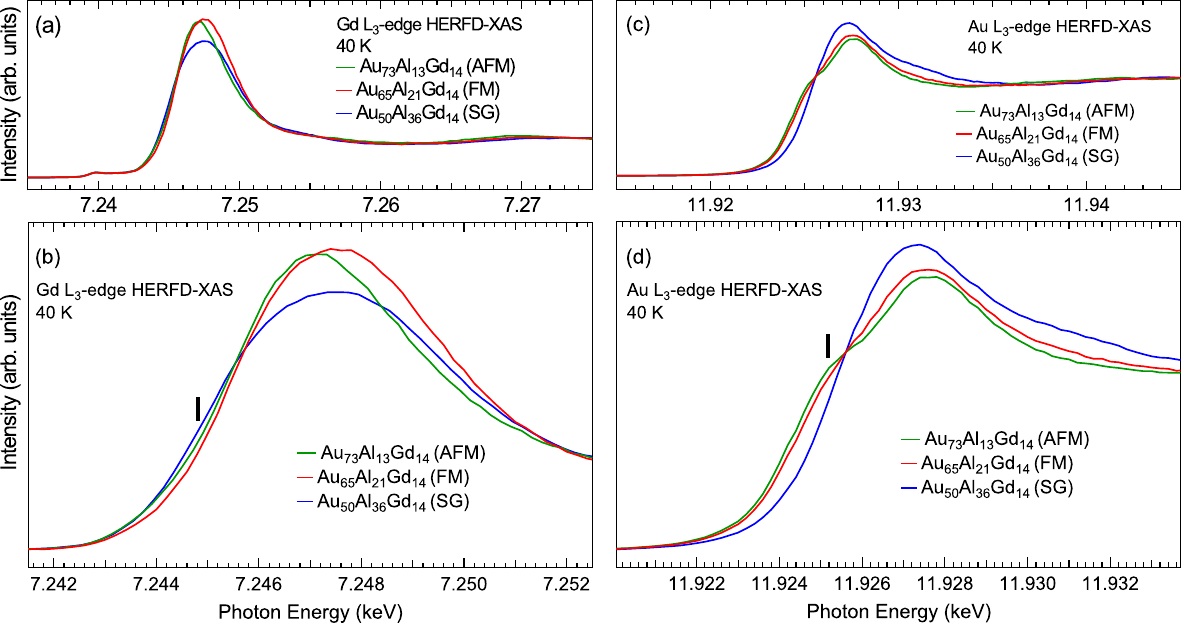}
	\caption{(a) Gd $L_3$-edge HERFD-XAS spectra of $\rm{Au_{73}}$$\rm{Al_{13}}$$\rm{Gd_{14}}$ (AFM), $\rm{Au_{65}}$$\rm{Al_{21}}$$\rm{Gd_{14}}$ (FM), and $\rm{Au_{50}}$$\rm{Al_{36}}$$\rm{Gd_{14}}$ (SG). These spectra are normalized by the spectral weights in high ($>$ 7.313 keV) and low ($<$ 7.241 keV) photon energy regions. (b) Enlarged view of the Gd $L_3$-edge HERFD-XAS spectra of $\rm{Au_{73}}$$\rm{Al_{13}}$$\rm{Gd_{14}}$ (AFM), $\rm{Au_{65}}$$\rm{Al_{21}}$$\rm{Gd_{14}}$ (FM), and $\rm{Au_{50}}$$\rm{Al_{36}}$$\rm{Gd_{14}}$ (SG). The solid line indicates the tail structure around 7.245 keV. (c) Au $L_3$-edge HERFD-XAS spectra of $\rm{Au_{73}}$$\rm{Al_{13}}$$\rm{Gd_{14}}$ (AFM), $\rm{Au_{65}}$$\rm{Al_{21}}$$\rm{Gd_{14}}$ (FM), and $\rm{Au_{50}}$$\rm{Al_{36}}$$\rm{Gd_{14}}$ (SG). These spectra are normalized using the spectral weights in the high- ($>$ 11.996 keV) and low- ($<$ 11.918 keV) photon-energy regions. (d) Expanded Au $L_3$-edge HERFD-XAS spectra for $\rm{Au_{73}}\rm{Al_{13}}\rm{Gd_{14}}$ (AFM), $\rm{Au_{65}}\rm{Al_{21}}\rm{Gd_{14}}$ (FM), and $\rm{Au_{50}}\rm{Al_{36}}\rm{Gd_{14}}$ (SG). The solid line shows the hump structure around 11.925 keV.}
	\label{Fig8}
	\end{center}
\end{figure*}

\begin{table}[b]
\caption{Relative photoionization cross sections per electron $\sigma$ for Au 5$d$, Au 6$s$, Al 3$s$, Al 3$p$, Gd 4$f$, Gd 5$d$, and Gd 6$s$  orbitals at the photon energy of 7 keV \cite{ac}.}
\label{sigma}
\centering
\scalebox{0.98}{
 \begin{tabular}{cccccccc}\hline
 Orbital\hspace{1mm} &Au 5$d$\hspace{1mm} &Au 6$s$\hspace{1mm} &Al 3$s$\hspace{1mm} &Al 3$p$\hspace{1mm} &Gd 4$f$\hspace{1mm} &Gd 5$d$\hspace{1mm} &Gd 6$s$\\ \hline
$\sigma$\hspace{1mm} &1\hspace{1mm} &0.294\hspace{1mm} &0.091\hspace{1mm} &0.003\hspace{1mm} &0.104\hspace{1mm} &0.192\hspace{1mm} &0.137\\ \hline
\label{Table1}
\end{tabular}}
\end{table}

\subsection{Valence-band HAXPES}
As discussed above, the variation of the Gd 5$d$ states is revealed by the Gd 3$d$-edge RPES. This modulation would be related to the magnetic properties of the Au-Al-Gd ACs because the Gd 5$d$ band would be hybridized with Au and/or Al valence bands for the intersite exchange interaction. Therefore, it is essential to probe the valence electronic states of Au and/or Al sites. We have also performed the valence-band HAXPES on the Au-Al-Gd ACs. Figure  \ref{Fig6}(a) displays the valence-band HAXPES spectra of the Au-Al-Gd ACs with that of the Au metal as a reference. As shown in Table \ref{Table1}, these spectra dominantly reflect the Au 5$d$ electronic states due to the largest photoionization cross section of the Au 5$d$ orbital at the photon energy of 7.2 keV \cite{ac}. For the Au metal, a prominent structure is observed from 2  to 8 eV, which is ascribed to the Au 5$d$ states. Distinct structures, which have different spectral shapes and bandwidths from those of the Au metal, are confirmed from $\sim$3 to $\sim$8 eV in the valence-band spectra of the Au-Al-Gd ACs. These structures originate from the Au 5$d$ bands \cite{ad}.\par

In the comparison of the valence-band HAXPES spectra of the Au-Al-Gd ACs, the peak binding energy (indicated by the solid lines) of the Au 5$d$ bands varies among the Au-Al-Gd ACs. This shift is plotted as a function of the $e/a$ ratio in Fig.~\ref{Fig6}(b). The peak binding energy is shifted toward the higher binding energy side with a larger value of the $e/a$ ratio. This behavior can be interpreted as a chemical shift due to the increasing number of conduction electrons. On the other hand, the Au 5$d$ bandwidth (indicated by the dashed lines) is also changed among these spectra in Fig.~\ref{Fig6}(a). To clearly show this bandwidth change, the Au 5$d$ bandwidth is also plotted as a function of the $e/a$ ratio in Fig. \ref{Fig6}(b). The bandwidth is decreased with a larger value of the $e/a$ ratio. This change in the bandwidth, which is not simply explained by the chemical shift, implies the systematic modulation of the electronic structure as a function of $e/a$ ratio or Au concentration. The decrease in Au concentration might lead to the 5$d$-band narrowing caused by the reduction of the effective Au 5$d$-5$d$ hybridizations. In the previous study, the reduction in the Au 5$d$ bandwidth compared with that of the Au metal has also been observed in the valence-band HAXPES spectra of the Au-Ga-Ce ACs \cite{ad}. This bandwidth reduction could be widely observed in the Au-containing rare-earth-based ACs. The clear difference in the Au 5$d$ electronic states is observed around the high binding energy region from $E_F$ in the Au-Al-Gd ACs.\par

Figure \ref{Fig7}(a) displays the enlarged view of the valence-band HAXPES spectra of the Au-Al-Gd ACs near $E_F$. The spectrum of $\rm{Au_{50}Al_{36}Gd_{14}}$ (SG) shows the broad peak structure around 1 eV. In contrast, the two peak structures are observed in the spectra of $\rm{Au_{73}Al_{13}Gd_{14}}$ (AFM) and $\rm{Au_{65}}$$\rm{Al_{21}Gd_{14}}$ (FM). Interestingly, the two peaks are centered at around 0.2  eV and 0.9 eV in $\rm{Au_{65}}$$\rm{Al_{21}Gd_{14}}$ (FM), while they are located at around 0.4 eV and 1.2 eV in $\rm{Au_{73}Al_{13}Gd_{14}}$ (AFM). The difference in the Au 5$d$ electronic states near $E_F$ is observed not only between the spin-glass phase sample and the samples with long-range magnetic order, but also between the antiferromagnet and ferromagnet samples in the paramagnetic phase. \par

For a direct comparison between the Gd 5$d$ and Au 5$d$ electronic states, we show the valence-band HAXPES spectra together with the difference spectra from the Gd 3$d$-edge RPES in Fig.~\ref{Fig7}(b), representing the Gd 5$d$ contribution. For $\rm{Au_{73}Al_{13}Gd_{14}}$ (AFM) and $\rm{Au_{65}}$$\rm{Al_{21}Gd_{14}}$ (FM), the peak binding energies near $E_F$ are close to each other in the difference spectra and the valence-band HAXPES spectra. This similarity indicates that the  Gd 5$d$ orbitals are hybridized with the Au 5$d$ orbitals (Gd 5$d$$-$Au 5$d$ hybridization) at around $E_F$. In contrast, for $\rm{Au_{50}Al_{36}Gd_{14}}$ (SG), the peak binding energy in the valence-band HAXPES spectrum is located on the higher binding energy side compared with that in the difference spectra, which suggests that the effective Gd 5$d$$–$Au 5$d$ hybridization strength in $\rm{Au_{50}Al_{36}Gd_{14}}$ (SG) is weaker than that in $\rm{Au_{73}Al_{13}Gd_{14}}$ (AFM) and $\rm{Au_{65}Al_{21}Gd_{14}}$ (FM). \par

\subsection{Gd and Au $L_3$-edge HERFD-XAS}
To support the results of the Gd 3$d$-edge RPES and valence-band HAXPES discussed above, the unoccupied Gd and Au 5$d$ electronic states are further investigated by the Gd and Au $L_3$-edge HERFD-XAS, respectively, which is a complementary method for probing orbital-resolved valence electronic states. Figure \ref{Fig8}(a) presents the Gd $L_3$-edge HERFD-XAS spectra of the Au-Al-Gd ACs.  In all spectra, there is a main peak structure around 7.247 keV, which is attributed to the 2$p$$-$5$d$ transition. A shoulder structure is also observed around 7.240 keV in all spectra, which originates from the 2$p$-4$f$ quadrupole transition \cite{ae}. The main peak around 7.247 keV reflects the unoccupied Gd 5$d$ electronic states. The spectral intensity of the main peak in $\rm{Au_{73}Al_{13}Gd_{14}}$ (AFM) and $\rm{Au_{65}}$$\rm{Al_{21}Gd_{14}}$ (FM) is higher than that in $\rm{Au_{50}Al_{36}Gd_{14}}$ (SG). This tendency is discussed later in conjunction with the Au $L_3$-edge HERFD-XAS results. To discuss the unoccupied Gd 5$d$ electronic states, we display the enlarged view of the Gd $L_3$-edge HERFD-XAS spectra around the main peak in Fig. \ref{Fig8}(b). The spectral intensity around 7.245 keV is mutually different among the Au-Al-Gd ACs. Therefore, the different Gd 5$d$ electronic states on the unoccupied side are confirmed by the Gd $L_3$-edge HERFD-XAS.\par

Figure \ref{Fig8}(c) shows the Au $L_3$-edge HERFD-XAS spectra of the Au-Al-Gd ACs. The main peak due to the 2$p$$-$5$d$ transition is observed around 11.928 keV in all spectra, reflecting the unoccupied Au 5$d$ electronic states. In contrast to the Gd $L_3$-edge HERFD-XAS spectra, the spectral intensity of the main peak in $\rm{Au_{50}Al_{36}Gd_{14}}$ (SG) is higher than that in $\rm{Au_{73}Al_{13}Gd_{14}}$ (AFM) and $\rm{Au_{65}}$$\rm{Al_{21}Gd_{14}}$ (FM). Figure \ref{Fig8}(d) displays the enlarged view around the main peak. For $\rm{Au_{73}Al_{13}Gd_{14}}$ (AFM) and $\rm{Au_{65}}$$\rm{Al_{21}Gd_{14}}$ (FM), the shoulder structure is observed around 11.925 keV. On the other hand, this hump structure is suppressed or smeared out in $\rm{Au_{50}Al_{36}Gd_{14}}$ (SG). Thus, the variations in the unoccupied Au 5$d$ electronic states are observed by the Au $L_3$-edge HERFD-XAS.\par

By comparing the Gd and Au $L_3$-edge HERFD-XAS results shown in Fig.~\ref{Fig8}, we find that the tendency of the main-peak intensity is opposite between these two edges. This shows that the behavior of the main-peak intensity cannot be simply explained by the change of the occupied valence electronic states with the $e/a$ ratio. Importantly, for $\rm{Au_{50}Al_{36}Gd_{14}}$ (SG), the main peak at the Gd $L_3$ edge is lower and broader than that for $\rm{Au_{73}Al_{13}Gd_{14}}$ (AFM) and $\rm{Au_{65}}$$\rm{Al_{21}Gd_{14}}$ (FM). In contrast, the main peak of $\rm{Au_{50}Al_{36}Gd_{14}}$ (SG) at the Au $L_3$ edge is higher and narrower than that of $\rm{Au_{73}Al_{13}Gd_{14}}$ (AFM) and $\rm{Au_{65}}$$\rm{Al_{21}Gd_{14}}$ (FM). With the strong Gd 5$d$$–$Au 5$d$ hybridization, the HERFD-XAS spectra are expected to show a similar behavior at both edges due to the hybridized bands \cite{q}. Based on that, the opposite tendency between the Gd and Au $L_3$-edge HERFD-XAS can be considered a fingerprint of the weak effective Gd 5$d$$–$Au 5$d$ hybridization in $\rm{Au_{50}Al_{36}Gd_{14}}$ (SG). It should be noted that the HERFD-XAS spectral shapes do not directly reflect the unoccupied 5$d$ partial density of states due to the interaction between the 5$d$ orbital and the core holes in the HERFD-XAS process.\par

\subsection{Discussion}
Our results show that the valence electronic states vary in the Au-Al-Gd ACs with the different magnetic ground states, while the Gd 4$f$ electrons are localized in all of them. Furthermore, we reveal the Gd 5$d$$–$Au 5$d$ hybridization around $E_F$, of which the effective strength varies among the samples. Although the origin of the valence-electronic-state modulation around $E_F$ cannot be clarified solely from the present results, the observed changes in valence electronic structures would be important for the difference in the magnetic properties of the Au-Al-Gd ACs through the RKKY interaction.\par

To understand the relationship between the observed valence electronic states and the magnetic properties, we now focus on the origin of the intersite exchange interaction between the localized Gd 4$f$ electrons.  It is known that there is an intra-atomic exchange interaction between the localized Gd 4$f$ electrons and the Gd 5$d$ states \cite{ax}. Combining the observed Gd 5$d$$-$Au 5$d$ hybridization with this intra-atomic Gd 4$f$$-$Gd 5$d$ exchange interaction, we can envisage a local pathway for the intersite exchange interaction between localized Gd 4$f$ electrons at different sites, in which neighboring Gd sites are connected through the Gd 5$d$$-$Au 5$d$$-$Gd 5$d$ local path. This picture indicates that the Gd 5$d$$-$Au 5$d$ hybridization would play a significant role in the intersite exchange interaction in the Au-Al-Gd ACs.\par

The previous discussion on the magnetic properties of the Au-Al-Gd ACs does not consider the orbital character of conduction electrons and hybridization between conduction electrons, such as Gd 5$d$$-$Au 5$d$ hybridization \cite{j}. Thus, we suggest that the Gd 5$d$$-$Au 5$d$ hybridization should also be taken into account as an additional important key for the magnetic properties of these systems. Note that the magnetic properties are not determined solely by the Gd 5$d$$-$Au 5$d$ hybridization. Disordered Au/Al sites \cite{k} and geometrical frustration are important for the SG ground states. In addition, we do not exclude the importance of the $k_F$ shift, which has been taken into account in the previous study \cite{j}. Angle-resolved photoemission spectroscopy \cite{ay}, which directly provides information about $k_F$, would be useful for further discussion.\par

The effective hybridization strength in $\rm{Au_{50}Al_{36}Gd_{14}}$ (SG) is relatively suppressed compared with that in $\rm{Au_{73}Al_{13}Gd_{14}}$ (AFM) and $\rm{Au_{65}}$$\rm{Al_{21}Gd_{14}}$ (FM). This suggests that the Gd 5$d$$-$Au 5$d$ hybridization strength would depend on the $e/a$ ratio and/or the Au concentration. Previous structural analyses of the Au-Al-Gd ACs \cite{k} have shown that the occupancy of Au on mixed Au/Al sites decreases with decreasing Au concentration, which corresponds to an increase in the $e/a$ ratio. Thus, the number of Au neighbors around a Gd site decreases with increasing $e/a$ ratio from the AFM phase to the SG phase in Fig.~\ref{Fig1}(a). This could contribute to the suppression of the Gd 5$d$$-$Au 5$d$ hybridization strength in the SG phase. In addition, the relative contribution of Al 3$sp$ states to the valence band could also be important for the Gd 5$d$$-$Au 5$d$ hybridization strength. Because the Gd 5$d$ and Au 5$d$ orbitals share the same $d$-orbital symmetry, their spatial overlap is expected to be larger than that between the Gd 5$d$ and Al 3$sp$ orbitals. Since the increasing Al concentration corresponds to an increase in the $e/a$ ratio, this orbital character difference could further reduce the Gd 5$d$$-$Au 5$d$ hybridization strength in the SG phase.\par

\section{Summary}
In summary, we have performed the soft and hard x-ray electron spectroscopies on the Au-Al-Gd ACs. The Gd 4$f$ electrons are localized in all samples. In contrast, the variation of the Gd 5$d$ and Au 5$d$ electronic states is clarified by both photoemission spectroscopy and HERFD-XAS. From the comparison of the difference spectra from the Gd 3$d$-edge RPES and valence-band HAXPES spectra, we reveal that the Gd 5$d$ states are hybridized with the Au 5$d$ states around $E_F$, which provides a microscopic picture of the local pathway for the intersite exchange interaction between the localized Gd 4$f$ electrons. The effective Gd 5$d$$–$Au 5$d$ hybridization strength is weaker in $\rm{Au_{50}Al_{36}Gd_{14}}$ (SG) than that in $\rm{Au_{73}Al_{13}Gd_{14}}$ (AFM) and $\rm{Au_{65}Al_{21}Gd_{14}}$ (FM). These findings show that the orbital character of the valence electronic states around $E_F$, such as the Gd 5$d$$-$Au 5$d$ hybridization, should be taken into account for the discussion of the magnetic properties of the Au-Al-Gd ACs.\par

\section{Acknowledgments}
The soft x-ray measurements at SPring-8 BL17SU were performed under the approval of RIKEN (Proposals No. 20230077 and 20240072), whereas the hard x-ray measurements at SPring-8 BL09XU (Proposals No. 2023B1930 and 2024A1645) and BL39XU (Proposals No. 2023A1872, 2023A1324, 2024A1963, 2024A1965, 2024A1972, and 2024B1720) were performed under the approval of JASRI. This work was financially supported by a Grant-in-Aid  for Innovative Areas (JP20H05271 and JP22H04594), a Grant-in-Aid for Transformative Research (JP23H04867), a Grant-in-Aid for Scientific Research (JP22K03527, JP23K21828, and JP24K03202) from JSPS and MEXT, Japan, and CREST  (JPMJCR22O3) from JST, Japan. G. Nozue was supported by the University of Osaka fellowship program of Super Hierarchical Material Science Program and the  JSPS Research Fellowships for Young Scientists.

\begin{figure}[t]
	\begin{center}
	\includegraphics[keepaspectratio,,scale=0.63]{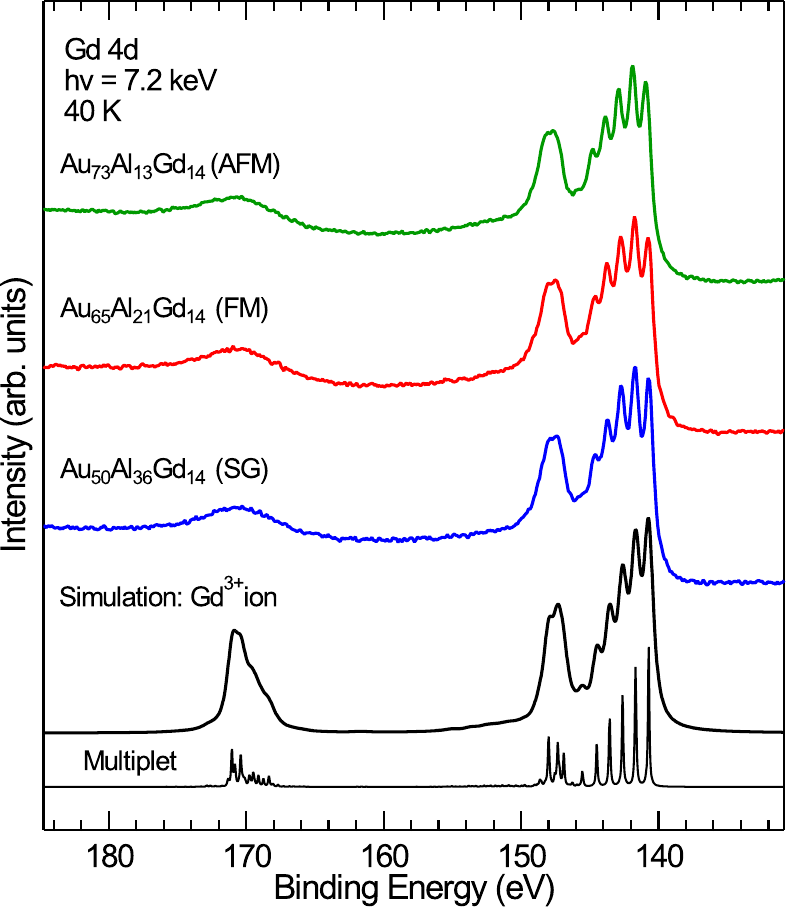}
	\caption{Gd 4$d$ core-level HAXPES spectra of $\rm{Au_{73}}$$\rm{Al_{13}}$$\rm{Gd_{14}}$ (AFM), $\rm{Au_{65}}$$\rm{Al_{21}}$$\rm{Gd_{14}}$ (FM), and $\rm{Au_{50}}$$\rm{Al_{36}}$$\rm{Gd_{14}}$ (SG) in comparison with the simulated spectrum for the Gd$^{3+}$ ion and the 4$d^9$4$f^7$ final-state multiplet structures.}
	\label{Fig9}
	\end{center}
\end{figure}

\begin{figure}[t]
	\begin{center}
	\includegraphics[keepaspectratio,,scale=0.65]{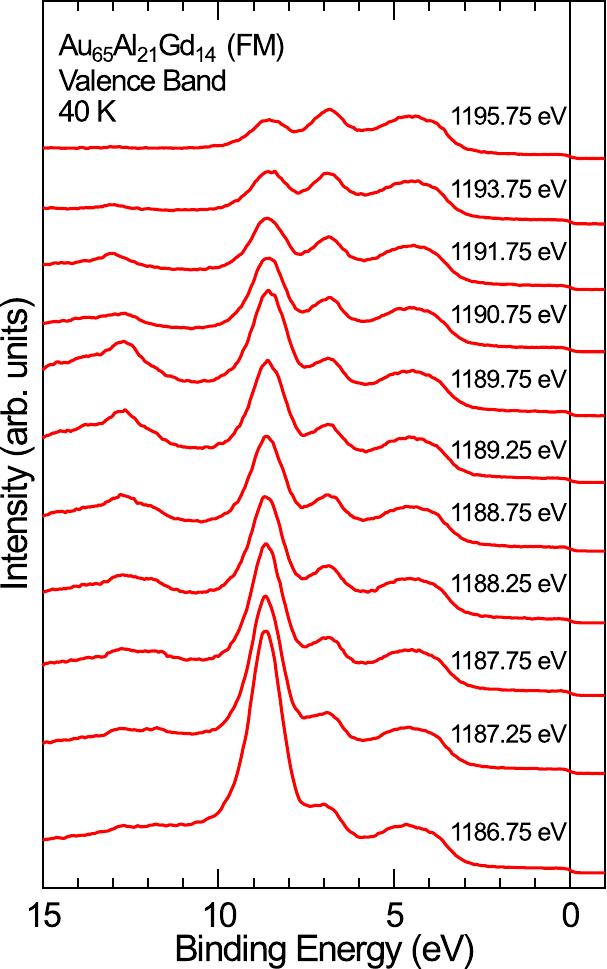}
	\caption{Gd 3$d$-edge RPES spectra of $\rm{Au_{65}}\rm{Al_{21}}\rm{Gd_{14}}$ (FM) across the photon energy of 1189 eV.}
	\label{Fig10}
	\end{center}
\end{figure}

\section{Appendices}
\subsection{Gd 4$d$ core-level HAXPES}
To cross-check the localized Gd 4$f$ electronic states, we have also performed the Gd 4$d$ core-level HAXPES on the Au-Al-Gd ACs. Figure \ref{Fig9} shows the Gd 4$d$ core-level HAXPES spectra of $\rm{Au_{73}Al_{13}Gd_{14}}$ (AFM), $\rm{Au_{65}Al_{21}Gd_{14}}$ (FM), and $\rm{Au_{50}Al_{36}Gd_{14}}$ (SG), where the linear backgrounds have been subtracted from the raw spectra of all Au-Al-Gd ACs. The overall spectral shape is mutually similar in these Gd 4$d$ core-level HAXPES spectra. The calculated spectrum for the Gd$^{3+}$ ion and the simulated 4$d^9$4$f^7$ final-state multiplet structures are also displayed in Fig.~\ref{Fig9}. Here, these simulations were calculated by the XTLS ver.~9.0 program \cite{s}, whereas the 4$d$-4$f$ Slater integrals and 4$d$ spin-orbit coupling were calculated by Cowan's code \cite{t}. In addition, the 4$d$-4$f$ Slater integrals and 4$d$ spin-orbit coupling were reduced to 74\% and 99\% to reproduce the experimental spectral shape, respectively. Regarding the 4$f$-4$f$ Slater integrals and the 4$f$ spin-orbit coupling, we have used the same parameters for the Gd 3$d$ core-level calculated spectrum. The simulated spectrum explains the line shape of the experimental Gd 4$d$ core-level HAXPES spectra. In particular, the several peaks in the experimental spectra are well explained by the 4$d^9$4$f^7$ final-state multiplet structures \cite{af}, indicating that the localized Gd$^{3+}$ state is dominant in the Au-Al-Gd ACs.

\subsection{Gd 3$d$-edge RPES spectra of $\rm{Au_{65}Al_{21}Gd_{14}}$ (FM)}
Figure \ref{Fig10} shows the Gd 3$d$-edge RPES spectra of $\rm{Au_{65}Al_{21}Gd_{14}}$ (FM) around the photon energy of 1189 eV. The peak structure around the binding energy of 8.7 eV is ascribed to the 4$f$$^6$ final-state multiplet structures. The broad two-peak structures from $\sim$3 to $\sim$8 eV originate from the Au 5$d$ bands. In addition to these structures, the hump structures are observed around the binding energy of 13 eV in the spectra at $h$$\nu$$\sim$1189.25 eV, which can be understood as the effect of the spin-flip transition in the Gd 3$d$-edge RPES \cite{w,x}.

\subsection{Temperature-dependent Gd and Au $L_3$-edge HERFD-XAS}
\begin{figure}
	\begin{center}
	\includegraphics[keepaspectratio,,scale=0.4]{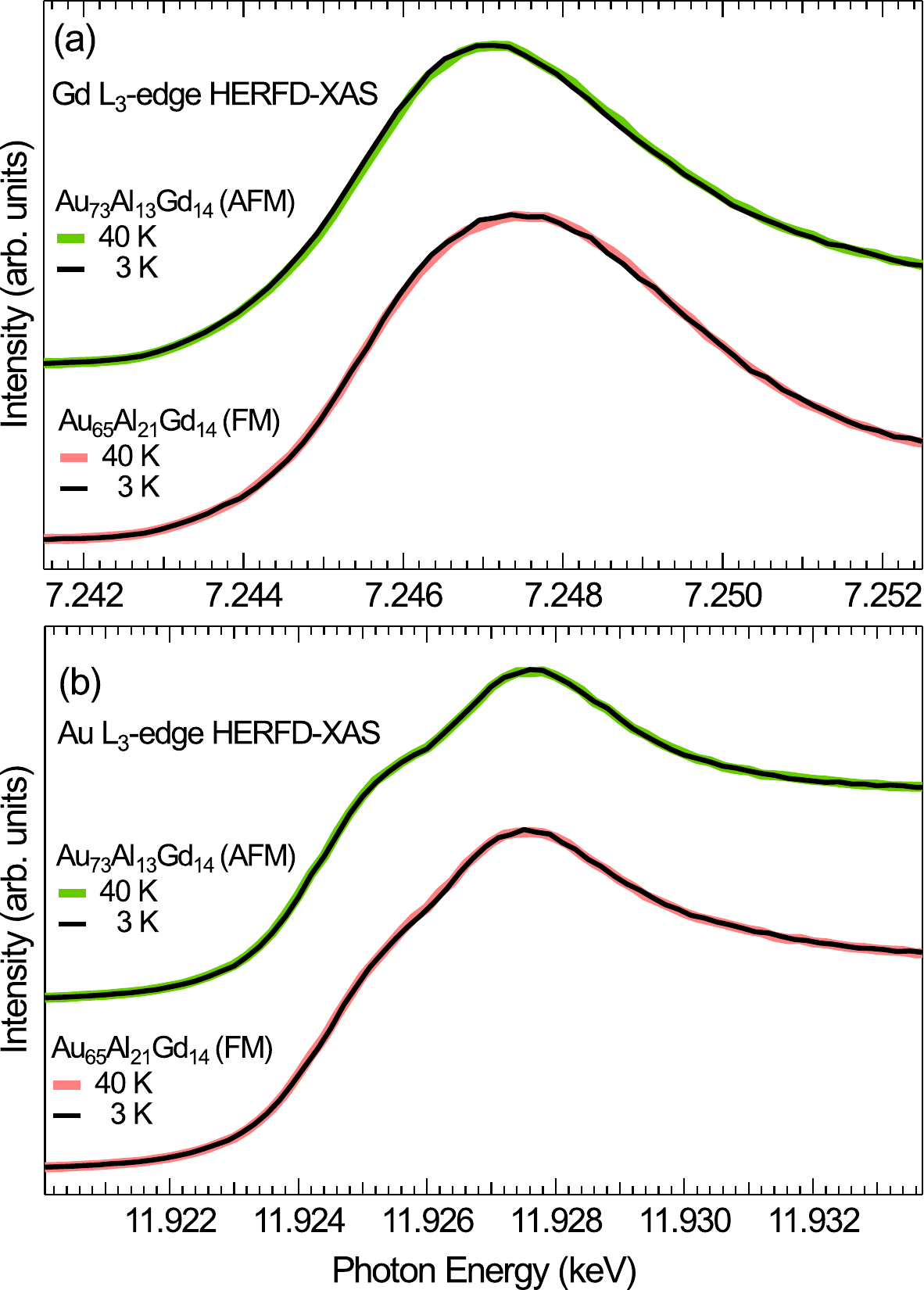}
	\caption{(a), (b) Temperature dependence of the Gd $L_3$-edge and the Au $L_3$-edge HERFD-XAS spectra of  $\rm{Au_{73}}$$\rm{Al_{13}}$$\rm{Gd_{14}}$ (AFM) and $\rm{Au_{65}}$$\rm{Al_{21}}$$\rm{Gd_{14}}$ (FM), respectively.}
	\label{Fig11}
	\end{center}
\end{figure}

For the discussion on the temperature dependence of valence electronic states, we have carried out the Gd and Au $L_3$-edge HERFD-XAS on $\rm{Au_{73}Al_{13}Gd_{14}}$ (AFM) and $\rm{Au_{65}Al_{21}Gd_{14}}$ (FM) at 40 K and 3 K, below the N\'{e}el temperature and the Curie temperature. Figures \ref{Fig11}(a) and (b) present the temperature dependence of the Gd and Au $L_3$-edge HERFD-XAS spectra of $\rm{Au_{73}Al_{13}Gd_{14}}$ (AFM) and $\rm{Au_{65}Al_{21}Gd_{14}}$ (FM), respectively. The line shape at 40 K (paramagnetic phase) is similar to that at 3 K (antiferromagnetic and ferromagnetic phase) in both $\rm{Au_{73}Al_{13}Gd_{14}}$ (AFM) and $\rm{Au_{65}Al_{21}Gd_{14}}$ (FM), suggesting that the Gd 5$d$ and Au 5$d$ electronic states are similar between the paramagnetic state and the magnetic ground states in the energy scale of a few to ten eV. These results allow us to discuss the relationship between the valence electronic states and the magnetic properties based on the spectra measured in the paramagnetic phase.

\end{document}